\documentclass[12pt]{article}
\usepackage{amssymb,amsmath,amsthm,amscd,latexsym}
\usepackage{amsfonts}
\usepackage[mathscr]{eucal}
\usepackage{color,hyperref}
\usepackage{graphicx}
\usepackage{multirow}
\usepackage{authblk}

\newtheorem{thm}{Theorem}[section]

\def\1v{\mathbf{1}}
\def\0v{\mathbf{0}}

\begin{document}

\title{New Self-Dual Codes of length 68 from a $2 \times 2$ block matrix Construction and Group Rings}

\author[$\dagger$]{M. Bortos}
\author[$\star$]{J. Gildea}
\author[$\ddag$]{A. Kaya}
\author[$\star$]{A. Korban}
\author[$\dagger$]{A. Tylyshchak}

\affil[$\dagger$]{Department of Algebra, Uzhgorod National University, Uzhgorod, Ukraine}
\affil[$\star$]{Department of Mathematical and Physical Sciences, University of Chester, UK}
\affil[$\ddag$]{Sampoerna University, 12780, Jakarta, Indonesia}

\maketitle

\begin{abstract}
Many generator matrices for constructing extremal binary self-dual codes of different lengths have the form $G=(I_n \ | \ A),$ where $I_n$ is the $n \times n$ identity matrix and $A$ is the $n \times n$ matrix fully determined by the first row. In this work, we define a generator matrix in which $A$ is a block matrix, where the blocks come from group rings and also, $A$ is not fully determined by the elements appearing in the first row. By applying our construction over $\mathbb{F}_2+u\mathbb{F}_2$ and by employing the extension method for codes, we were able to construct new extremal binary self-dual codes of length 68. Additionally, by employing a generalised neighbour method to the codes obtained, we were able to construct many new binary self-dual $[68,34,12]$-codes with the rare parameters $\gamma=7,8$ and $9$ in $W_{68,2}.$ In particular, we find 92 new binary self-dual $[68,34,12]$-codes.
\end{abstract}


\section{Introduction}

There has been a great interest and focus on constructing binary self-dual codes of different lengths with new weight enumerators. Researches have employed different methods and tools to search for these codes and as a result, many new codes have been discovered. For example, a classical approach is to use a generator matrix of the form $G=(I_n \ | \ A),$ where $I_n$ is the $n \times n$ identity matrix and $A$ is a special matrix, for example, a circulant matrix. Please see \cite{AI, AII, AIII} for example. Another well known method is to search for self-dual codes over rings, and then to consider their binary images to produce extremal binary self-dual codes with new weight enumerators. Please see \cite{AIV, AV, AVI, AVII} for examples. The well known extension and neighbour methods \cite{AVII} can also be used to search for new extremal binary self-dual codes. Recently, the neighbour method has been generalised in \cite{AVIII} so that the $k^{th}$ range of neighbours can be constructed leading to finding new self-dual codes. This generalised method turned out to be a very powerful tool as many new codes can be obtained by considering the family of neighbours of just one code. Please see \cite{AVIII} for details.

In this work, we present a generator matrix of the form $M_{\sigma}=(I_n \ | \ A),$ where $A$ is a block matrix obtained from group rings. Similar generator matrices can be found in \cite{AII, AIII}. Our construction differs from the other, known in the literature generator matrices, since we make sure that the block matrix $A$ is not fully determined by the blocks appearing in the first row- this has been the case in the generator matrices known in the literature \cite{AII, AIII}. By defining the block matrix $A$ in this way, we force the search field to be greater than in the standard generator matrices which leads us to finding self-dual codes that are not attainable from other techniques. By applying our construction over the ring $\mathbb{F}_2+u\mathbb{F}_2,$ by considering the binary images and by employing the well known extension method, the neighbour technique and its generalisation, we find new extremal binary self-dual codes of length 68. In particular, we find such codes with new weight enumerators for the rare parameters $\gamma=7,8$ and $9.$

The rest of the paper is organised as follows. In Section~2, we give preliminary definitions and results on self-dual codes, the alphabets to be used, group rings, the well known extension method, the neighbour technique and its generalization from \cite{AVIII}. In Section~3, we introduce the new construction and give theoretical results. We show when our construction produces self-dual codes. In Section~4, we apply our main construction over the alphabet $\mathbb{F}_2+u\mathbb{F}_2$ and consider their binary images to find extremal binary self-dual codes of length 64. We next apply the extension method to search for binary self-dual codes with parameters $[68,34,12].$ Lastly, we employ the neighbour technique and its generalisation to find extremal binary self-dual codes of length 68 with weight enumerators that were not known in the literature before. In particular, we find codes with new weight enumerator with the rare parameters $\gamma=7,8$ and $9.$ Altogether, we find $98$ new extremal binary self-dual codes of length 68. We also tabulate all the numerical results in this section. W finish with concluding remarks and directions for possible future research.

\section{Preliminaries}

\subsection{Self-Dual Codes, the Ring $\mathbb{F}_2+u\mathbb{F}_2,$ Group Rings and the Well Known Extension and Neighbor Methods}

We begin by recalling the standard definitions from coding theory. A code $C$
of length $n$ over a Frobenius ring $R$ is a subset of $R^n$. If the code is
a submodule of $R^n$ then we say that the code is linear. Elements of the
code $C$ are called codewords of $C$. Let $\mathbf{x}=(x_1,x_2,\dots,x_n)$
and $\mathbf{y}=(y_1,y_2,\dots,y_n)$ be two elements of $R^n.$ The duality
is understood in terms of the Euclidean inner product, namely:
\begin{equation*}
\langle \mathbf{x},\mathbf{y} \rangle_E=\sum x_iy_i.
\end{equation*}
The dual $C^{\bot}$ of the code $C$ is defined as
\begin{equation*}
C^{\bot}=\{\mathbf{x} \in R^n \ | \ \langle \mathbf{x},\mathbf{y}
\rangle_E=0 \ \text{for all} \ \mathbf{y} \in C\}.
\end{equation*}
We say that $C$ is self-orthogonal if $C \subseteq C^\perp$ and is self-dual
if $C=C^{\bot}.$

An upper bound on the minimum Hamming distance of a binary self-dual code
was given in \cite{AIX}. Specifically, let $d_{I}(n)$ and $d_{II}(n)$ be the
minimum distance of a Type~I and Type~II binary code of length $n$,
respectively. Then
\begin{equation*}
d_{II}(n) \leq 4\lfloor \frac{n}{24} \rfloor+4
\end{equation*}
and
\begin{equation*}
d_{I}(n)\leq
\begin{cases}
\begin{matrix}
4\lfloor \frac{n}{24} \rfloor+4 \ \ \ if \ n \not\equiv 22 \pmod{24} \\
4\lfloor \frac{n}{24} \rfloor+6 \ \ \ if \ n \equiv 22 \pmod{24}.%
\end{matrix}%
\end{cases}%
\end{equation*}

Self-dual codes meeting these bounds are called \textsl{extremal}.
Throughout the text, we obtain extremal binary codes of different lengths.
Self-dual codes which are the best possible for a given set of parameters is
said to be optimal. Extremal codes are necessarily optimal but optimal codes
are not necessarily extremal.

\subsection{The ring $\mathbb{F}_2+u\mathbb{F}_2$}

In this section, we recall some theory on self-dual codes over $\mathbb{F}%
_2+u\mathbb{F}_2.$ We refer to \cite{AVI} where Type II, Type IV, self-dual
codes and cyclic codes over $\mathbb{F}_2+u\mathbb{F}_2$ have been studied.

The ring $\mathbb{F}_2+u\mathbb{F}_2$ is a ring of characteristic 2 with 4
elements with the restriction $u^2=0.$ It is defined as
\begin{equation*}
\mathbb{F}_2+u\mathbb{F}_2=\{a+bu \ | \ a,b \in \mathbb{F}_2, u^2=0\},
\end{equation*}
and it is easily seen that $\mathbb{F}_2+u\mathbb{F}_2 \cong \mathbb{F}%
_2[x]/(x^2).$ A linear code $C$ of length $n$ over the ring $\mathbb{F}_2+u%
\mathbb{F}_2$ is an $\mathbb{F}_2+u\mathbb{F}_2$-submodule of $(\mathbb{F}%
_2+u\mathbb{F}_2)^n.$ The elements of $\mathbb{F}_2+u\mathbb{F}_2$ are $%
0,1,u,1+u$ and their Lee weights are defined as $0,1,2,1$ respectively. The
Hamming $(d_H)$ and Lee $(d_L)$ distance between $n$ tuples is then defined
as the sum of the Hamming and Lee weights of the difference of the
components of these tuples respectively. The smallest positive Hamming and
Lee distance of a code $C$ is denoted by $d_H(C)$ and $d_L(C)$ respectively.

A Gray map $\phi$ is defined as
\begin{equation*}
\phi : (\mathbb{F}_2+u\mathbb{F}_2) \rightarrow \mathbb{F}_2^{2n},
\end{equation*}
\begin{equation*}
\phi(\overline{a}+\overline{b}u)=(\overline{b},\overline{a}+\overline{b}),
\end{equation*}
where $\overline{a}, \overline{b} \in \mathbb{F}_2^n.$ The map is a distance
preserving isometry from $((\mathbb{F}_2+u\mathbb{F}_2)^n, d_L)$ to $(%
\mathbb{F}_2^{2n},d_H),$ where $d_L$ and $d_H$ denote the Lee and Hamming
distance in $(\mathbb{F}_2+u\mathbb{F}_2)^n$ and $\mathbb{F}_2^{2n}$
respectively. This means that if $C$ is a linear code over $\mathbb{F}_2+u%
\mathbb{F}_2$ with parameters $[n,2^k,d]$ ($2^k$ is the number of the
codewords), then $\phi(C)$ is a binary linear code of parameters $[2n,k,d].$
The following theorem is a natural result of the Gray map.

\begin{thm}
\label{thm.2.2.1} If $C$ is a self-dual code over $\mathbb{F}_2+u\mathbb{F}_2
$ of length $n,$ then $\phi(C)$ is a self-dual binary code of length $2n.$
\end{thm}

We can also define a natural projection from $\mathbb{F}_2+u\mathbb{F}_2$ to
$\mathbb{F}_2$ as follows:
\begin{equation*}
\mu : \mathbb{F}_2+u\mathbb{F}_2 \rightarrow \mathbb{F}_2,
\end{equation*}
\begin{equation*}
\mu(a+bu)=a.
\end{equation*}
If $D=\mu(C)$ for some linear code $C$ over $\mathbb{F}_2+u\mathbb{F}_2,$ we
say that $D$ is a projection of $C$ into $\mathbb{F}_2,$ and that $C$ is a
lift of $D$ into $\mathbb{F}_2+u\mathbb{F}_2.$ It is clear that the
projection of a self-orthogonal code is self-orthogonal, but the projection
of a self-dual code need not be self-dual. We now state two well known results which we apply later in our work when searching for self-dual codes.

\begin{thm}
\label{thm.2.2.2} Suppose that $C$ is a self-dual code over $\mathbb{F}_2+u%
\mathbb{F}_2$ of length $2n,$ generated by the matrix $[I_n|A],$ where $I_n$
is the $n \times n$ identity matrix. Then $\mu(C)$ is a self-dual binary
code of length $2n.$
\end{thm}

\begin{thm}
\label{thm.2.2.3} Suppose $C$ is a linear code over $\mathbb{F}_2+u\mathbb{F}%
_2$ and that $C^{\prime }=\mu(C),$ is its projection to $\mathbb{F}_2.$ With
$d$ and $d^{\prime }$ representing the minimum Lee and Hamming distances of $%
C$ and $C^{\prime }$ respectively, we have that $d \leq 2d^{\prime }.$
\end{thm}

For the computational results in later sections, we are going to use the
following extension method to obtain codes of length $n+2.$

\begin{thm}\label{extensionthm} (\cite{AVII}) Let $\mathcal{C}$ be a self-dual code of
length $n$ over a commutative\ Frobenius ring with identity $R$ and $G=(r_i)$
be a $k \times n$ generator matrix for $\mathcal{C}$, where $r_i $ is the
i-th row of $G $, $1\leq i \leq k.$ Let $c$ be a unit in $R$ such that $%
c^2=-1$ and $X$ be a vector in $S^n$ with $\langle X,X \rangle=-1.$ Let $%
y_i=\langle r_i, X \rangle $ for $1 \leq i \leq k.$ The following matrix

\begin{equation*}
\begin{bmatrix}
\begin{tabular}{cc|c}
$1$ & $0$ & $X$ \\ \hline
$y_1$ & $cy_1$ & $r_1$ \\
$\vdots$ & $\vdots$ & $\vdots$ \\
$y_k$ & $cy_k$ & $r_k$%
\end{tabular}%
\end{bmatrix}%
,
\end{equation*}
generates a self-dual code $\mathcal{D}$ over $R$ of length $n+2.$
\end{thm}

We will also apply the neighbor method and its generalization to search for new extremal binary
self-dual codes from codes obtained directly from our main construction or from
the described above, extension method. Two self-dual binary codes of length $%
2n$ are said to be neighbours of each other if their intersection has
dimension $n-1$. Let $x\in {\mathbb{F}}_{2}^{2n} \setminus \mathcal{C}$ then
$\mathcal{D}=\left\langle \left\langle x\right\rangle ^{\bot }\cap \mathcal{C%
},x\right\rangle $ is a neighbour of $\mathcal{C}$.

Recently in \cite{AVIII}, the neighbor method has been extended and the following formula for constructing the $k^{th}$-range neighbour codes was provided:

$$\mathcal{N}_{(i+1)}=\left\langle \left\langle x_i \right\rangle^{\bot} \cap \mathcal{N}_{(i)}, x_i \right\rangle,$$

where $\mathcal{N}_{(i+1)}$ is the neighbour of $\mathcal{N}_{(i)}$ and $x_i \in \mathbb{F}_2^{2n}-\mathcal{N}_{(i)}.$

\subsection{Group Rings}

We first give the definitions of some special matrices which we use later in our work. A circulant matrix is one where each row is shifted one element to the right relative to the preceding row. We label the circulant matrix as $A=circ(\alpha_1,\alpha_2\dots , \alpha_n),$ where $\alpha_i$ are ring elements. A block-circulant matrix is one where each row contains blocks which are square matrices. The rows of the block matrix are defined by shifting one block to the right relative to the preceding row. We label the block-circulant matrix as $CIRC(A_1,A_2,\dots A_n),$ where $A_i$ are $k \times k$ matrices over the ring $R.$ The transpose of a matrix $A,$ denoted by $A^T,$ is a matrix whose rows are the columns of $A,$ i.e., $A^T_{ij}=A_{ji}.$ We finish this section by giving the necessary definitions for group rings and by recalling the map that sends a group ring element $v \in RG$ to a $n \times n$ matrix over $R.$.

While group rings can be given for infinite rings and infinite groups, we
are only concerned with group rings where both the ring and the group are
finite. Let $G$ be a finite group of order $n$, then the group ring $RG$
consists of $\sum_{i=1}^n \alpha_i g_i$, $\alpha_i \in R$, $g_i \in G.$

Addition in the group ring is done by coordinate addition, namely
\begin{equation}
\sum_{i=1}^n \alpha_i g_i +\sum_{i=1}^n \beta_i g_i =\sum_{i=1}^n (\alpha_i
+ \beta_i)g_i.
\end{equation}
The product of two elements in a group ring is given by
\begin{equation}
\left(\sum_{i=1}^n \alpha_i g_i \right)\left(\sum_{j=1}^n \beta_j g_j
\right)= \sum_{i,j} \alpha_i \beta_j g_i g_j.
\end{equation}
It follows that the coefficient of $g_i$ in the product is $\sum_{g_i
g_j=g_k} \alpha_i \beta_j.$

The following construction of a matrix was first given for codes over fields
by Hurley in \cite{AX}. It was extended to Frobenius rings in \cite{AXI}. Let $%
R$ be a finite commutative Frobenius ring and let $G=\{g_1,g_2,\dots,g_n\}$
be a group of order $n$. Let $v = \sum_{i=1}^n \alpha_{g_i}   \in RG.$  Define the 
matrix $\sigma(v) \in M_n(R)$ to be $\sigma(v)=(\alpha_{g_i^{-1} g_j})$ where 
$i,j \in \{1,2,\cdots,n\}$.

We note that the elements $g_1^{-1},
g_2^{-1}, \dots, g_n^{-1}$ are the elements of the group $G$ in a some given
order. We will now describe $\sigma(v)$ for the following group rings $RG$
where $G \in \{C_8 \ \text{and} \ D_{8}\}$.

\begin{enumerate}
\item[(i)] Let $C_n=\langle x \ | \ x^n=1 \rangle$ and set our listing of $%
C_n$ to be $\{1,x,x^2,\dots,x^{n-1}\}$. Let $v=\displaystyle{\sum_{i=0}^{n-1}%
} \alpha_i x^i \in RC_n$ where $\alpha_i \in R$, where $R$ is a ring, then
\begin{equation}
\sigma(v)=circ(\alpha_0, \alpha_1,\dots, \alpha_{n-1}).
\end{equation}

\item[(ii)] Let $G= \langle x,y \ | \ x^4=y^2=1, x^y=x^{-1} \rangle \cong
D_{8}.$ If $v=\sum_{i=0}^3 \alpha_{i+1}x^i+\alpha_{i+5}x^iy \in RD_{8},$
then
\begin{equation}  \label{RD_{16}}
\sigma(v)=%
\begin{pmatrix}
A & B \\
B^T & A^T%
\end{pmatrix}%
\end{equation}
where $A=circ(\alpha_1,\alpha_2,\alpha_3,\alpha_4),$ $B=circ(\alpha_5,\alpha_6,%
\alpha_7,\alpha_8)$ and $\alpha_i \in R.$
\end{enumerate}

We also recall the canonical involution $* : RG \rightarrow RG$ on a group
ring $RG$ is given by $v^*=\sum_{g}\alpha_g g^{-1},$ for $v=\sum_g \alpha_g
g \in RG.$ An important connection between $v^*$ and $v$ appears when we
take their images under the $\sigma$ map:
\begin{equation}
\sigma(v^*)=\sigma(v)^T.
\end{equation}
If $v$ satisfies $vv^*=1,$ then we say that $v$ is a unitary unit in $RG.$

\section{Main Construction}

In this section, we present the main construction in this work. Let $v_i \in RG$ where $R$ is a finite commutative Frobenius ring of characteristics 2 with $1\leq i \leq 3$ and $G$ be a finite group of order $n.$ Define the following matrix:
\begin{equation}
M_{\sigma}=\begin{bmatrix}
\begin{tabular}{c|cc}
\multirow{2}{*}{$I_{2n}$} & $\sigma(v_1)$ & $\sigma(v_2)$ \\
                          & $\sigma(v_2)$ & $\sigma(v_3)$
\end{tabular}
\end{bmatrix}.
\end{equation}

Let $C_{\sigma}$ be a code that is generated by the matrix $M_{\sigma}.$ Then, the code $C_{\sigma}$ has length $4n.$ We now state the main result of this work.

\begin{thm}\label{MainTheorem}
Let $R$ be a finite commutative Frobenius ring of characteristics 2 and let $G$ be a finite group of order $n.$ Then $C_{\sigma}$ is a self-dual code of length $4n$ if and only if
\begin{equation}
\sigma(v_1v_1^*+v_2v_2^*)=I_n,
\end{equation}
\begin{equation}
v_1v_2^*+v_2v_3^*=0,
\end{equation}
\begin{equation}
v_2v_1^*+v_3v_2^*=0,
\end{equation}
\begin{equation}
\sigma(v_2v_2^*+v_3v_3^*)=I_n.
\end{equation}
\begin{proof}
The code generated will be self-dual if and only if $M_{\sigma}M_{\sigma}^T$ is the zero matrix over $R.$ Since $M_{\sigma}$ is of the form $(I \ | \ A),$ where
$$A=\begin{pmatrix}
\sigma(v_1)&\sigma(v_2)\\
\sigma(v_2)&\sigma(v_3)
\end{pmatrix},$$
it is enough to show that $AA^T=I_{2n}.$ Now,
$$AA^T=\begin{pmatrix}
\sigma(v_1)&\sigma(v_2)\\
\sigma(v_2)&\sigma(v_3)
\end{pmatrix}\begin{pmatrix}
\sigma(v_1^*)&\sigma(v_2^*)\\
\sigma(v_2^*)&\sigma(v_3^*)
\end{pmatrix}=\begin{pmatrix}
\sigma(v_1v_1^*+v_2v_2^*)&\sigma(v_1v_2^*+v_2v_3^*)\\
\sigma(v_2v_1^*+v_3v_2^*)&\sigma(v_2v_2^*+v_3v_3^*)
\end{pmatrix}.$$
Clearly, $AA^T=I_{2n}$ iff $\sigma(v_1v_1^*+v_2v_2^*)=I_n,$ $v_1v_2^*+v_2v_3^*=0,$ $v_2v_1^*+v_3v_2^*=0$ and $\sigma(v_2v_2^*+v_3v_3^*)=I_n.$
\end{proof}
\end{thm}
We have immediately that the code has free rank $2n$ by construction. It is also obvious from the construction that the search field of $M_{\sigma}$ when searching for self-dual codes over $R$ is $|R|^{3n}.$ We note that the blocks $\sigma(v_1),$ $\sigma(v_2)$ and $\sigma(v_3)$ are independent of each other which makes our construction different than the usual constructions. A typical generator matrix of a self-dual code has the form $(I|A)$ where $A$ is a special matrix in which the rows are simply permutations of the first row, for example, a circulant or reverse-circulant matrices. In our construction, the block $\sigma(v_3)$ makes the difference, namely, the rows in the matrix $M_{\sigma}$ are not anymore permutations of the elements of the first row because of $\sigma(v_3).$ If $v_1=v_3$ then the matrix $M_{\sigma}$ consists of rows which are permutations of the elements in the first row, but because $v_3$ is independent of $v_1$ and $v_2$ this is not the case.

\section{Computational Results}

In this section, we employ our main construction, the extension method, the neighbor technique and its generalization to search for extremal self-dual binary codes of length 68. Namely, we apply the main construction over the field $\mathbb{F}_2$ to search for self-dual codes of length 32. We then lift these to codes over the ring $\mathbb{F}_2+u\mathbb{F}_2$ whose binary images are the extremal self-dual binary codes of length 64. Next we consider the extensions of the codes of length 64 to obtain extremal self-dual binary codes with parameters $[68,34,12].$ Finally, we apply the neighbor technique and its generalization to the codes of length 68 to find many codes of that length with weight enumerators not known in the literature before. In particular we find many new codes with the rare parameters $\gamma=7,8$ and $9.$ Before we tabulate the results, we recall the weight enumerators of extremal self-dual binary codes with parameters $[64,32,12]$ and $[68,34,12].$

There are two possibilities for the weight enumerators of extremal singly-even $[64,32,12]_2$ codes (\cite{AXII}):

\begin{equation*}
W_{64,1}=1+(1312+16\beta)y^{12}+(22016-64\beta)y^{14}+\dots, \ 14 \leq \beta
\leq 284,
\end{equation*}
\begin{equation*}
W_{64,2}=1+(1312+16\beta)y^{12}+(23040-64\beta)y^{14}+ \dots, \ 0 \leq \beta
\leq 277.
\end{equation*}
Recently, many new codes are constructed for both weight enumerators in \cite{AXIII}, \cite{AXIV} and \cite{AXV}. With the most updated information, the existence
of codes is known for $\beta =$14, 16, 18, 19, 20, 22, 24, 25, 26, 28, 29,
30, 32, 34, 35, 36, 38, 39, 44, 46, 49, 53, 54, 58, 59, 60, 64 and $74$ in $%
W_{64,1}$ and for $\beta =0,\dots ,$40, 41, 42, 44, 45, 46, 47, 48, 49, 50,
51, 52, 54, 55, 56, 57, 58, 60, 62, 64, 69, 72, 80, 88, 96, 104, 108, 112,
114, 118, 120 and $184$ in $W_{64,2}.$

The weight
enumerator of a self-dual $[68,34,12]_2$ code is in one of the following
forms by \cite{AXVI, AXVII}:
\begin{equation*}
W_{68,1}=1+(442+4\beta)y^{12}+(10864-8\beta)y^{14}+\dots,
\end{equation*}
\begin{equation*}
W_{68,2}=1+(442+4\beta)y^{12}+(14960-8\beta-256\gamma)y^{14}+\dots \ ,
\end{equation*}
where $\beta$ and $\gamma$ are parameters and $0 \leq \gamma \leq 9.$ The
first examples of codes with a $\gamma =7$ in $W_{68,2}$ are constructed in
\cite{AXVIII}. The first examples of codes with $\gamma=8,9$ in $W_{68,2}$ are constructed in \cite{AVIII}. Together with these the existence of the codes in $W_{68,2}$
is known for the following parameters (see \cite{AIII, AXIX, AXX, AXXI, AXVIII, AXIII}):\newline

$%
\begin{array}{l}
\gamma =0,\ \beta \in \{2m|m=0,7,11,14,17,20,21,\dots
,99,100,102,105,110,119,136,165\};\ \text{or} \\
\qquad \beta \in \{2m+1|m=3,5,8,10,15,16,17,19,20,\dots ,82,87,91,\dots,99,101,104,110,\\
\qquad 115\};
\\
\gamma =1,\ \beta \in \{2m|m=19,22,\dots ,99,108\};\ \text{or}\\
\qquad \beta \in
\{2m+1|m=24,\dots ,85,94,100,101,106,108,116\}; \\
\gamma =2,\ \beta \in \{2m|m=29,\dots ,100,103,104\};\ \text{or}\

\beta \in\{2m+1|m=32,\dots ,81,84,85,86\}; \\
\gamma =3,\ \beta \in \{2m|m=39,\dots ,92,94,95,97,98,101,102\};\ \text{or}
\\
\qquad \beta \in \{2m+1|m=38,39,40,42,43,\dots ,77,79,80,81,83,87,88,89,96\}; \\
\gamma =4,\ \beta \in \{2m|m=43,46,\dots ,58,60,\dots ,93,97,98,100\};\
\text{or} \\
\qquad \beta \in \{2m+1|m=48,\dots ,55,57,58,60,61,62,64,68,\ldots
,72,74,78,79,80,83,84,\\
\qquad 85,89,95\}; \\
\gamma =5\ \beta \in \{m|m=113,116,\dots ,153,158,\dots
,169,182,187,189,191,193,195,198,\\
\qquad 200,202,211\}; \\
\gamma =6\ \text{with}\ \beta \in \{2m|m=69,77,78,79,81,88,91,93,94,95,97,\dots,103\}; \\
\qquad \text{or}\ \beta \in \{2m+1|m=87,\dots,100,103\};\\
\gamma =7\ \text{with}\ \beta \in \{2m|m=49,56,63,70,77,83,\dots,99,105,106,112,119,126,133,147\};\\
\qquad \text{or}\ \beta \in \{2m+1|m=52,59,66,73,80,81,84,\dots,99,101,108,115,122,129,136\};\\
\gamma =8\ \text{with}\ \beta \in \{2m|m=90,\dots,110\};\ \text{or}\ \beta \in \{2m+1|m=90,\dots,110\};\\
\gamma =9\ \text{with}\ \beta \in \{2m|m=93,\dots,105,107,\dots,115\};\ \text{or}\ \beta \in \{2m+1|m=93,94,96,97,99,\\
\qquad \dots,112\};\\
\end{array}%
$\newline

All the upcoming computational results were obtained by performing the searches using MAGMA (\cite{AXXII}).

\subsection{The group $C_8$}

We first consider the main construction with the cyclic group $C_8$ to search for binary self-dual codes with parameters $[32,16,6 \ \text{or} \ 8].$ We only list these codes which then lead to us finding new extremal binary self-dual codes of length 68.

\begin{table}[h!]
\caption{Codes of length 32 via Theorem \ref{MainTheorem} with the cyclic group $C_8$}\centering
\begin{tabular}{cccccc}
\hline
Code  & $v_1$               & $v_2$               & $v_3$               & $|Aut(C)|$  & Type          \\ \hline
$C_1$ & $(0,0,0,0,0,1,1,1)$ & $(0,0,0,0,0,1,0,1)$ & $(0,1,0,1,0,0,1,0)$ & $2^93^25$ & $[32,16,6]_I$ \\ \hline
$C_2$ & $(0,0,0,0,1,1,1,1)$ & $(0,0,1,1,0,1,1,1)$ & $(0,0,0,1,1,1,1,0)$ & $2^5$     & $[32,16,6]_I$ \\ \hline
\end{tabular}
\end{table}

We now consider the $R_1$ lifts of the codes from Table 1. The codes obtained have binary images of extremal binary self-dual codes with parameters $64,32,12.$ As before, we only list these codes that we later use to obtain new extremal binary self-dual codes of length 68.

\begin{table}[h!]\caption{Codes of length 64 from $R_1$ lifts of $C_1$ and $C_2$}
\resizebox{0.8\textwidth}{!}{\begin{minipage}{\textwidth}
\centering
\begin{tabular}{ccccccc}
\hline
Code  &       & $v_1$                 & $v_2$                   & $v_3$                     & $|Aut(C)|$ & $W_{64,2}$ \\ \hline
$I_1$ & $C_2$ & $(u,0,0,u,1,1,u+1,1)$ & $(0,0,1,u+1,0,u+1,1,1)$ & $(0,u,u,u+1,u+1,u+1,1,0)$ & $2^5$    & $\beta=0$  \\ \hline
$I_2$ & $C_1$ & $(0,u,0,0,0,1,1,u+1)$ & $(0,u,0,u,u,1,0,1)$     & $(0,1,0,u+1,0,0,1,u)$     & $2^7$    & $\beta=80$ \\ \hline
\end{tabular}
\end{minipage}}
\end{table}

\subsection{The group $D_8$}

Now we consider the main construction and the dihedral group $D_8$ to search for binary self-dual codes with parameters $[32,16,6 \ \text{or} \ 8].$

\begin{table}[h!]
\caption{Codes of length 32 via Theorem \ref{MainTheorem} with the dihedral group $D_8$}\centering
\begin{tabular}{cccccc}
\hline
Code  & $v_1$               & $v_2$               & $v_3$               & $|Aut(C)|$  & Type          \\ \hline
$C_3$ & $(0,0,0,1,0,0,1,1)$ & $(0,0,1,1,0,1,0,1)$ & $(1,1,0,1,0,1,1,0)$ & $2^33$     & $[32,16,6]_I$ \\ \hline
\end{tabular}
\end{table}

$R_1$ lifts:

\begin{table}[h!]\caption{Codes of length 64 from $R_1$ lifts of $C_3$}
\resizebox{0.8\textwidth}{!}{\begin{minipage}{\textwidth}
\centering
\begin{tabular}{ccccccc}
\hline
Code  &       & $v_1$                 & $v_2$                   & $v_3$                     & $|Aut(C)|$ & $W_{64,2}$ \\ \hline
$I_3$ & $C_3$ & $(0,u,u,1,0,0,1,1)$ & $(0,0,1,u+1,u,1,0,1)$ & $(u+1,u+1,0,u+1,0,1,u+1,0)$ & $2^43$    & $\beta=64$  \\ \hline
\end{tabular}
\end{minipage}}
\end{table}

\subsection{Extremal Binary Self-Dual Codes of length 68 via Extensions}

We now apply Theorem~\ref{extensionthm} to the codes obtained in Tables 2 and 4. As a result, we obtain codes whose binary images are the extremal binary self-dual codes of length 68. The order of the automorphism group of all the codes obtained in the table below is 2.

\begin{table}[h!]\caption{New codes of length 68 from Theorem \ref{extensionthm}}
\resizebox{0.85\textwidth}{!}{\begin{minipage}{\textwidth}
\centering
\begin{tabular}{cccccc}
\hline
$C_{68,i}$ & Code & $(x_{17},x_{18},\dots ,x_{32})$ & $c$ & $\gamma $ & $%
\beta \ in\ W_{64,2}$ \\ \hline
$C_{68,1}$ & $I_{3}$ & $%
(u,u,0,1,3,u,u,u,u,3,u,1,0,0,u,3,u,1,0,1,1,3,3,3,u,0,3,u,u,1,0,0)$ & $1$ & $%
\boldsymbol{0}$ & $\boldsymbol{181}$ \\ \hline
$C_{68,2}$ & $I_{3}$ & $%
(0,3,0,1,3,1,3,1,1,1,3,1,u,0,u,0,u,0,u,3,0,0,1,1,1,1,u,u,1,3,u,u)$ & $3$ & $%
\boldsymbol{1}$ & $\boldsymbol{185}$ \\ \hline
$C_{68,3}$ & $I_{1}$ & $%
(0,1,1,u,u,3,u,1,3,3,1,0,0,3,3,u,1,3,3,u,u,3,0,u,3,u,3,u,1,3,0,0)$ & $3$ & $%
\boldsymbol{2}$ & $\boldsymbol{54}$ \\ \hline
$C_{68,4}$ & $I_{2}$ & $%
(u,3,1,3,0,0,3,u,0,3,0,u,u,u,0,u,3,1,0,3,0,3,u,1,1,1,1,u,0,3,0,1)$ & $1$ & $%
\boldsymbol{2}$ & $\boldsymbol{202}$ \\ \hline
$C_{68,5}$ & $I_{3}$ & $%
(u,u,u,3,0,0,1,u,1,u,1,3,u,0,0,3,0,1,u,3,0,1,0,3,1,1,0,3,u,3,0,1)$ & $1$ & $%
\boldsymbol{3}$ & $\boldsymbol{179}$ \\ \hline
$C_{68,6}$ & $I_{3}$ & $%
(u,0,0,1,1,0,1,0,3,3,u,0,1,0,3,3,1,0,3,0,3,3,1,u,1,u,1,u,3,0,1,u)$ & $3$ & $%
\boldsymbol{3}$ & $\boldsymbol{189}$ \\ \hline
$C_{68,7}$ & $I_{3}$ & $%
(0,3,0,1,u,3,u,3,0,1,0,3,0,3,0,0,1,u,u,1,0,u,1,0,u,0,u,1,3,0,1,u)$ & $3$ & $%
\boldsymbol{3}$ & $\boldsymbol{198}$ \\ \hline
\end{tabular}
\end{minipage}}
\end{table}

\subsection{Extremal Binary Self-Dual Codes of length 68 via Neighbours}

We now apply the $k^{th}$ range neighbour formula (mentioned earlier) to one of the codes obtained in Table 5, namely, the code $C_{68,4}.$

Let \noindent $\mathcal{N}_{(0)}$ be $\gamma=2$, $\beta=202$ (Table 5 ($C_{68,4}$)), then we obtain the following codes by applying the $k^{th}$ range formula:

\begin{table}[h!]\caption{$i^{th}$ neighbour of  $\mathcal{N}_{(0)}$}
\resizebox{0.7\textwidth}{!}{\begin{minipage}{\textwidth}
\centering
\begin{tabular}{|c|c|ccc|c|c|ccc|}
\hline
$i$ & $\mathcal{N}_{(i+1)}$   & $x_i$  & $\gamma$ & $\beta$ & $i$ & $\mathcal{N}_{(i+1)}$   & $x_i$  & $\gamma$ & $\beta$  \\ \hline \hline
$0$ & $\mathcal{N}_{(1)}$ &  $(1110000001001111010001001000010000)$  & $3$ & $180$   &
$1$ & $\mathcal{N}_{(2)}$ &  $(0110111111100111000010000110011111)$  & $4$ & $177$   \\ \hline
$2$ & $\mathcal{N}_{(3)}$ &  $(1111110110010011100101001000101111)$  & $5$ & $169$   &
$3$ & $\mathcal{N}_{(4)}$ &  $(0100000000110011110000010000011110)$  & $6$ & $191$   \\ \hline
$4$ & $\mathcal{N}_{(5)}$ &  $(0100000000001101110010001110000110)$  & $6$ & $199$   &
$5$ & $\mathcal{N}_{(6)}$ &  $(0000100000001100010011001110000111)$  & $7$ & $199$   \\ \hline
$6$ & $\mathcal{N}_{(7)}$ &  $(1011111101001111000101010111111010)$  & $\textbf{7}$ & $\textbf{209}$   &
$7$ & $\mathcal{N}_{(8)}$ &  $(1110011111110110000101111101111110)$  & $\textbf{7}$ & $\textbf{220}$   \\ \hline
$8$ & $\mathcal{N}_{(9)}$ &  $(1100101001011011001101000000111100)$  & $8$ & $212$   &
$9$ & $\mathcal{N}_{(10)}$ &  $(1110110100111111000010011111011000)$  & $\textbf{8}$ & $\textbf{226}$   \\ \hline
$10$ & $\mathcal{N}_{(11)}$ &  $(1011101011111010111010001000101000)$  & $\textbf{8}$ & $\textbf{233}$   &
$11$ & $\mathcal{N}_{(12)}$ &  $(1101100000000101110010111111001110)$  & $9$ & $213$   \\ \hline
$12$ & $\mathcal{N}_{(13)}$ &  $(0111110101110100100110100100000111)$  & $9$ & $222$   &
$13$ & $\mathcal{N}_{(14)}$ &  $(1100011000000000100101010101100010)$  & $\textbf{9}$ & $\textbf{229}$   \\ \hline
$14$ & $\mathcal{N}_{(15)}$ &  $(0100111000100000110010100011000100)$  & $\textbf{9}$ & $\textbf{235}$   &
$15$ & $\mathcal{N}_{(16)}$ &  $(0000111011111101001101111010001101)$  & $\textbf{9}$ & $\textbf{236}$   \\ \hline
$16$ & $\mathcal{N}_{(17)}$ &  $(0110010111000111101001101101101110)$  & $\textbf{9}$ & $\textbf{240}$   &
$17$ & $\mathcal{N}_{(18)}$ &  $(0001101000010111100111111110001001)$  & $\textbf{9}$ & $\textbf{243}$   \\ \hline
$18$ & $\mathcal{N}_{(19)}$ &  $(1111010011001111000010010001010001)$  & $\textbf{9}$ & $\textbf{247}$   &
$19$ & $\mathcal{N}_{(20)}$ &  $(0001000000100010001011101011110000)$  & $\textbf{8}$ & $\textbf{234}$   \\ \hline
$20$ & $\mathcal{N}_{(21)}$ &  $(0110110001101011101000111100110001)$  & $\textbf{8}$ & $\textbf{245}$   &
$21$ & $\mathcal{N}_{(22)}$ &  $(1000011100010110100110011011000011)$  & $\textbf{8}$ & $\textbf{250}$   \\ \hline
\end{tabular}
\end{minipage}}
\end{table}

We shall now separately consider the neighbours of $\mathcal{N}_{(7)}, \mathcal{N}_{(8)}, \mathcal{N}_{(10)}, \mathcal{N}_{(11)}, \mathcal{N}_{(12)}, \mathcal{N}_{(14)}, \mathcal{N}_{(15)},\\
\mathcal{N}_{(17)}, \mathcal{N}_{(18)}, \mathcal{N}_{(19)}, \mathcal{N}_{(20)}, \mathcal{N}_{(21)}$ and $\mathcal{N}_{(22)}.$

First, the neighbours of $\mathcal{N}_{(7)}.$

\begin{table}[h!]\caption{Neighbours of  $\mathcal{N}_{(7)}$}
\resizebox{0.7\textwidth}{!}{\begin{minipage}{\textwidth}
\centering
\begin{tabular}{|c|c|ccc|c|c|ccc|}
\hline
$\mathcal{N}_{(i)}$ & $\mathcal{M}_{i}$ & $(x_{35},x_{36},...,x_{68})$ & $\gamma$ & $\beta$ & $\mathcal{N}_{(i)}$ & $\mathcal{M}_{i}$ & $(x_{35},x_{36},...,x_{68})$ & $\gamma$ & $\beta$ \\ \hline
$7$ & $1$ & $(0001101011101000000010101100100000)$  & $\textbf{7}$ & $\textbf{200}$   &
$7$ & $2$ & $(1011110000101000000000000110110010)$  & $\textbf{7}$ & $\textbf{201}$   \\ \hline
$7$ & $3$ & $(1010110010100001100100011110001110)$  & $\textbf{7}$ & $\textbf{202}$   &
$7$ & $4$ & $(0001001110011000100100000010000111)$  & $\textbf{7}$ & $\textbf{204}$   \\ \hline
$7$ & $5$ & $(0110000000101001011011010010111000)$  & $\textbf{7}$ & $\textbf{205}$   &
$7$ & $6$ & $(1101010010010101110001000011001000)$  & $\textbf{7}$ & $\textbf{206}$   \\ \hline
$7$ & $7$ & $(0000101100001000100101011000010101)$  & $\textbf{7}$ & $\textbf{207}$   &
$7$ & $8$ & $(1101100101111100101110100100000011)$  & $\textbf{7}$ & $\textbf{212}$   \\ \hline
$7$ & $9$ & $(1111011000001111101001111011111111)$  & $\textbf{7}$ & $\textbf{214}$   &
 &  &   &  & \\ \hline
\end{tabular}
\end{minipage}}
\end{table}

The neighbours of $\mathcal{N}_{(8)}.$

\begin{table}[h!]\caption{Neighbours of  $\mathcal{N}_{(8)}$}
\resizebox{0.7\textwidth}{!}{\begin{minipage}{\textwidth}
\centering
\begin{tabular}{|c|c|ccc|c|c|ccc|}
\hline
$\mathcal{N}_{(i)}$ & $\mathcal{M}_{i}$ & $(x_{35},x_{36},...,x_{68})$ & $\gamma$ & $\beta$  & $\mathcal{N}_{(i)}$ & $\mathcal{M}_{i}$ & $(x_{35},x_{36},...,x_{68})$ & $\gamma$ & $\beta$ \\ \hline
$8$ & $10$ & $(0101101001001001100111000010001010)$  & $\textbf{6}$ & $\textbf{205}$   &
$8$ & $11$ & $(0000111000000101110110010000000101)$  & $\textbf{6}$ & $\textbf{211}$   \\ \hline
$8$ & $12$ & $(1000110101001001010000000111111011)$  & $\textbf{7}$ & $\textbf{208}$   &
$8$ & $13$ & $(1100001010000110010100101000001100)$  & $\textbf{7}$ & $\textbf{211}$   \\ \hline
$8$ & $14$ & $(0000011000010001001000011101100110)$  & $\textbf{7}$ & $\textbf{213}$   &
$8$ & $15$ & $(0011000110000110101101001101111011)$  & $\textbf{7}$ & $\textbf{215}$   \\ \hline
$8$ & $16$ & $(0100001111110010110100000101101010)$  & $\textbf{7}$ & $\textbf{216}$   &
$8$ & $17$ & $(1111101111010101000001000100011110)$  & $\textbf{7}$ & $\textbf{218}$   \\ \hline
\end{tabular}
\end{minipage}}
\end{table}

The neighbours of $\mathcal{N}_{(10)}.$

\begin{table}[h!]\caption{Neighbours of  $\mathcal{N}_{(10)}$}
\resizebox{0.7\textwidth}{!}{\begin{minipage}{\textwidth}
\centering
\begin{tabular}{|c|c|ccc|c|c|ccc|}
\hline
$\mathcal{N}_{(i)}$ & $\mathcal{M}_{i}$ & $(x_{35},x_{36},...,x_{68})$ & $\gamma$ & $\beta$ & $\mathcal{N}_{(i)}$ & $\mathcal{M}_{i}$ & $(x_{35},x_{36},...,x_{68})$ & $\gamma$ & $\beta$  \\ \hline
$10$ & $18$ & $(1000111101011101000010001111000100)$  & $\textbf{8}$ & $\textbf{222}$   &
$10$ & $19$ & $(1000001100101001110001001010110111)$  & $\textbf{8}$ & $\textbf{223}$   \\ \hline
$10$ & $20$ & $(0000100110010101011101101001100110)$  & $\textbf{8}$ & $\textbf{227}$   &
$10$ & $21$ & $(1011001101010011010111011000101010)$  & $\textbf{8}$ & $\textbf{229}$   \\ \hline
\end{tabular}
\end{minipage}}
\end{table}
\newpage
The neighbours of $\mathcal{N}_{(11)}.$

\begin{table}[h!]\caption{Neighbours of  $\mathcal{N}_{(11)}$}
\resizebox{0.7\textwidth}{!}{\begin{minipage}{\textwidth}
\centering
\begin{tabular}{|c|c|ccc|c|c|ccc|}
\hline
$\mathcal{N}_{(i)}$ & $\mathcal{M}_{i}$ & $(x_{35},x_{36},...,x_{68})$ & $\gamma$ & $\beta$ & $\mathcal{N}_{(i)}$ & $\mathcal{M}_{i}$ & $(x_{35},x_{36},...,x_{68})$ & $\gamma$ & $\beta$ \\ \hline
$11$ & $22$ & $(1010110110000101110101111100110110)$  & $\textbf{7}$ & $\textbf{221}$   &
$11$ & $23$ & $(0000001101000001110010110101100000)$  & $\textbf{7}$ & $\textbf{222}$   \\ \hline
$11$ & $24$ & $(1101010100100000111010001000010011)$  & $\textbf{8}$ & $\textbf{224}$   &
$11$ & $25$ & $(0000010011001000010100011111011111)$  & $\textbf{8}$ & $\textbf{225}$   \\ \hline
$11$ & $26$ & $(1110111110110010111011101101101110)$  & $\textbf{8}$ & $\textbf{228}$   &
$11$ & $27$ & $(1001100110100111000010100000100101)$  & $\textbf{8}$ & $\textbf{230}$   \\ \hline
$11$ & $28$ & $(0000110001111000001001000011101000)$  & $\textbf{8}$ & $\textbf{231}$   &
$11$ & $29$ & $(1001011111010011000001100001010000)$  & $\textbf{8}$ & $\textbf{232}$   \\ \hline
\end{tabular}
\end{minipage}}
\end{table}

The neighbours of $\mathcal{N}_{(12)}.$

\begin{table}[h!]\caption{Neighbours of  $\mathcal{N}_{(12)}$}
\resizebox{0.7\textwidth}{!}{\begin{minipage}{\textwidth}
\centering
\begin{tabular}{|c|c|ccc|c|c|ccc|}
\hline
$\mathcal{N}_{(i)}$ & $\mathcal{M}_{i}$ & $(x_{35},x_{36},...,x_{68})$ & $\gamma$ & $\beta$  & $\mathcal{N}_{(i)}$ & $\mathcal{M}_{i}$ & $(x_{35},x_{36},...,x_{68})$ & $\gamma$ & $\beta$\\ \hline
$12$ & $30$ & $(1000100110000001010101010100001001)$  & $\textbf{9}$ & $\textbf{191}$   &
$12$ & $31$ & $(0111010100101000000001100101011010)$  & $\textbf{9}$ & $\textbf{197}$   \\ \hline
$12$ & $32$ & $(1111100000101101001011110111000010)$  & $\textbf{9}$ & $\textbf{212}$   &
 &  &   &  &  \\ \hline
\end{tabular}
\end{minipage}}
\end{table}

The neighbours of $\mathcal{N}_{(14)}.$

\begin{table}[h!]\caption{Neighbour of  $\mathcal{N}_{(14)}$}
\resizebox{1\textwidth}{!}{\begin{minipage}{\textwidth}
\centering
\begin{tabular}{|c|c|ccc|c|c|ccc|}
\hline
$\mathcal{N}_{(i)}$ & $\mathcal{M}_{i}$ & $(x_{35},x_{36},...,x_{68})$ & $\gamma$ & $\beta$  \\ \hline
$14$ & $33$ & $(1011110001101000100111010000010000)$  & $\textbf{9}$ & $\textbf{227}$   \\ \hline
\end{tabular}
\end{minipage}}
\end{table}

The neighbours of $\mathcal{N}_{(15)}.$

\begin{table}[h!]\caption{Neighbours of  $\mathcal{N}_{(15)}$}
\resizebox{0.7\textwidth}{!}{\begin{minipage}{\textwidth}
\centering
\begin{tabular}{|c|c|ccc|c|c|ccc|}
\hline
$\mathcal{N}_{(i)}$ & $\mathcal{M}_{i}$ & $(x_{35},x_{36},...,x_{68})$ & $\gamma$ & $\beta$  & $\mathcal{N}_{(i)}$ & $\mathcal{M}_{i}$ & $(x_{35},x_{36},...,x_{68})$ & $\gamma$ & $\beta$\\ \hline
$15$ & $34$ & $(1011000011111110011101011000000101)$  & $\textbf{9}$ & $\textbf{231}$   &
$15$ & $35$ & $(1111110111110000010110000100010011)$  & $\textbf{9}$ & $\textbf{232}$   \\ \hline
$15$ & $36$ & $(0011000011010010100011010000111001)$  & $\textbf{9}$ & $\textbf{233}$   &
$15$ & $37$ & $(0000000000111100000000101100111101)$  & $\textbf{9}$ & $\textbf{234}$   \\ \hline
\end{tabular}
\end{minipage}}
\end{table}

The neighbours of $\mathcal{N}_{(17)}.$

\begin{table}[h!]\caption{Neighbours of  $\mathcal{N}_{(17)}$}
\resizebox{0.7\textwidth}{!}{\begin{minipage}{\textwidth}
\centering
\begin{tabular}{|c|c|ccc|c|c|ccc|}
\hline
$\mathcal{N}_{(i)}$ & $\mathcal{M}_{i}$ & $(x_{35},x_{36},...,x_{68})$ & $\gamma$ & $\beta$ & $\mathcal{N}_{(i)}$ & $\mathcal{M}_{i}$ & $(x_{35},x_{36},...,x_{68})$ & $\gamma$ & $\beta$ \\ \hline
$17$ & $38$ & $(0110000100000010110010110000110100)$  & $\textbf{9}$ & $\textbf{237}$  &
$17$ & $39$ & $(0011111001100000111100111101010010)$  & $\textbf{9}$ & $\textbf{238}$   \\ \hline
\end{tabular}
\end{minipage}}
\end{table}

The neighbours of $\mathcal{N}_{(18)}.$

\begin{table}[h!]\caption{Neighbours of  $\mathcal{N}_{(18)}$}
\resizebox{0.7\textwidth}{!}{\begin{minipage}{\textwidth}
\centering
\begin{tabular}{|c|c|ccc|c|c|ccc|}
\hline
$\mathcal{N}_{(i)}$ & $\mathcal{M}_{i}$ & $(x_{35},x_{36},...,x_{68})$ & $\gamma$ & $\beta$ & $\mathcal{N}_{(i)}$ & $\mathcal{M}_{i}$ & $(x_{35},x_{36},...,x_{68})$ & $\gamma$ & $\beta$ \\ \hline
$18$ & $40$ & $(0110010110000001001110111010011100)$  & $\textbf{9}$ & $\textbf{239}$   &
$18$ & $41$ & $(1111000010111111010100101000111101)$  & $\textbf{9}$ & $\textbf{241}$   \\ \hline
\end{tabular}
\end{minipage}}
\end{table}
\newpage
The neighbours of $\mathcal{N}_{(19)}.$

\begin{table}[h!]\caption{Neighbours of  $\mathcal{N}_{(19)}$}
\resizebox{0.7\textwidth}{!}{\begin{minipage}{\textwidth}
\centering
\begin{tabular}{|c|c|ccc|c|c|ccc|}
\hline
$\mathcal{N}_{(i)}$ & $\mathcal{M}_{i}$ & $(x_{35},x_{36},...,x_{68})$ & $\gamma$ & $\beta$ & $\mathcal{N}_{(i)}$ & $\mathcal{M}_{i}$ & $(x_{35},x_{36},...,x_{68})$ & $\gamma$ & $\beta$  \\ \hline
$19$ & $42$ & $(1011110110001100110101001011001010)$  & $\textbf{9}$ & $\textbf{242}$   &
$19$ & $43$ & $(0101010111011010111100000111011110)$  & $\textbf{9}$ & $\textbf{244}$   \\ \hline
$19$ & $44$ & $(1010110011000110001101001010010000)$  & $\textbf{9}$ & $\textbf{246}$   &
 &  &   &  & \\ \hline
\end{tabular}
\end{minipage}}
\end{table}

The neighbours of $\mathcal{N}_{(20)}.$

\begin{table}[h!]\caption{Neighbours of  $\mathcal{N}_{(20)}$}
\resizebox{0.7\textwidth}{!}{\begin{minipage}{\textwidth}
\centering
\begin{tabular}{|c|c|ccc|c|c|ccc|}
\hline
$\mathcal{N}_{(i)}$ & $\mathcal{M}_{i}$ & $(x_{35},x_{36},...,x_{68})$ & $\gamma$ & $\beta$ & $\mathcal{N}_{(i)}$ & $\mathcal{M}_{i}$ & $(x_{35},x_{36},...,x_{68})$ & $\gamma$ & $\beta$ \\ \hline
$20$ & $45$ & $(1111111110101111010101000110001101)$  & $\textbf{8}$ & $\textbf{236}$   &
$20$ & $46$ & $(1010000010110000100011110101111001)$  & $\textbf{8}$ & $\textbf{239}$   \\ \hline
\end{tabular}
\end{minipage}}
\end{table}

The neighbours of $\mathcal{N}_{(21)}.$

\begin{table}[h!]\caption{Neighbours of  $\mathcal{N}_{(21)}$}
\resizebox{0.7\textwidth}{!}{\begin{minipage}{\textwidth}
\centering
\begin{tabular}{|c|c|ccc|c|c|ccc|}
\hline
$\mathcal{N}_{(i)}$ & $\mathcal{M}_{i}$ & $(x_{35},x_{36},...,x_{68})$ & $\gamma$ & $\beta$ & $\mathcal{N}_{(i)}$ & $\mathcal{M}_{i}$ & $(x_{35},x_{36},...,x_{68})$ & $\gamma$ & $\beta$ \\ \hline
$21$ & $47$ & $(0100000110001011000001000101101010)$  & $\textbf{6}$ & $\textbf{208}$   &
$21$ & $48$ & $(1101111000000001010010000110110001)$  & $\textbf{6}$ & $\textbf{209}$   \\ \hline
$21$ & $49$ & $(1011101011101010010101111101000101)$  & $\textbf{6}$ & $\textbf{212}$   &
$21$ & $50$ & $(1111110111000100010010111011100000)$  & $\textbf{6}$ & $\textbf{214}$   \\ \hline
$21$ & $51$ & $(1011111101010010111011101111111100)$  & $\textbf{6}$ & $\textbf{215}$   &
$21$ & $52$ & $(0000000001100001001001100111011100)$  & $\textbf{6}$ & $\textbf{218}$   \\ \hline
$21$ & $53$ & $(1111011001110010100001101011011011)$  & $\textbf{6}$ & $\textbf{220}$   &
$21$ & $54$ & $(0100000001010101001001101001000011)$  & $\textbf{7}$ & $\textbf{219}$   \\ \hline
$21$ & $55$ & $(1100000000000001110100001001100111)$  & $\textbf{7}$ & $\textbf{223}$   &
$21$ & $56$ & $(0000001101000100110101111100001111)$  & $\textbf{7}$ & $\textbf{225}$   \\ \hline
$21$ & $57$ & $(1111011001111010111110100111110110)$  & $\textbf{7}$ & $\textbf{226}$   &
$21$ & $58$ & $(0010011000011000001000111001000101)$  & $\textbf{7}$ & $\textbf{227}$   \\ \hline
$21$ & $59$ & $(1001010101101111101110000000000011)$  & $\textbf{7}$ & $\textbf{230}$   &
$21$ & $60$ & $(1111110101100000100011001110100110)$  & $\textbf{8}$ & $\textbf{235}$   \\ \hline
$21$ & $61$ & $(0110000110110100100100101111100100)$  & $\textbf{8}$ & $\textbf{238}$   &
$21$ & $62$ & $(1010010010111110111001111011100010)$  & $\textbf{8}$ & $\textbf{240}$   \\ \hline
$21$ & $63$ & $(1101011100111011010011111101111110)$  & $\textbf{8}$ & $\textbf{241}$   &
 &  &   &  & \\ \hline
\end{tabular}
\end{minipage}}
\end{table}

Finally, the neighbours of $\mathcal{N}_{(22)}.$

\begin{table}[h!]\caption{Neighbours of  $\mathcal{N}_{(22)}$}
\resizebox{0.7\textwidth}{!}{\begin{minipage}{\textwidth}
\centering
\begin{tabular}{|c|c|ccc|c|c|ccc|}
\hline
$\mathcal{N}_{(i)}$ & $\mathcal{M}_{i}$ & $(x_{35},x_{36},...,x_{68})$ & $\gamma$ & $\beta$ & $\mathcal{N}_{(i)}$ & $\mathcal{M}_{i}$ & $(x_{35},x_{36},...,x_{68})$ & $\gamma$ & $\beta$  \\ \hline
$22$ & $63$ & $(0011111100111010011001010011100100)$  & $\textbf{5}$ & $\textbf{207}$   &
$22$ & $64$ & $(1011111100101111100110111111111101)$  & $\textbf{6}$ & $\textbf{213}$   \\ \hline
$22$ & $65$ & $(1001011101001100101011001000110100)$  & $\textbf{6}$ & $\textbf{217}$   &
$22$ & $66$ & $(0100111101000110110111101101111110)$  & $\textbf{6}$ & $\textbf{219}$   \\ \hline
$22$ & $68$ & $(1000010000111101010101110010010011)$  & $\textbf{7}$ & $\textbf{229}$   &
$22$ & $69$ & $(0100000001011101000011001111110011)$  & $\textbf{8}$ & $\textbf{237}$   \\ \hline
$22$ & $70$ & $(1111111101001111101100000010100000)$  & $\textbf{8}$ & $\textbf{242}$   &
$22$ & $71$ & $(0010000100001001100001001110111000)$  & $\textbf{8}$ & $\textbf{243}$   \\ \hline
$22$ & $72$ & $(1110110000001011011001101011011010)$  & $\textbf{8}$ & $\textbf{247}$   &
 &  &   &  &  \\ \hline
\end{tabular}
\end{minipage}}
\end{table}

Note that $|Aut(C)|=1$ for all the codes constructed.

\section{Conclusion}

In this paper, we presented a generator matrix of the form $G=(I_n \ | \ A),$ where $A$ is a block matrix where the blocks come from group rings and the rows of $A$ are not fully determined by permuting the entries of the first row. Together with our construction, extension and neighbours techniques, we were able to construct the following extremal binary self-dual codes with new weight enumerators in $W_{68,2}:$

\begin{equation*}
\begin{array}{l}
(\gamma =0,\ \beta =\{181\}), \\
(\gamma =1,\ \beta =\{185\}), \\
(\gamma =2,\ \beta =\{54,202\}), \\
(\gamma =3,\ \beta =\{179,189,198\}), \\
(\gamma =5,\ \beta =\{207\}), \\
(\gamma =6,\ \beta =\{205,208,209,211,212,213,214,215,217,218,219,220\}), \\
(\gamma =7,\ \beta =\{200,201,202,204,205,206,207,208,209,211,212,\\
\qquad \qquad \qquad 213,214,215,216,218,219,220,221,222,223,225,226,227,229,230\}), \\
(\gamma =8,\ \beta =\{222,223,224,225,226,227,228,229,230,231,232,233,234,235,236,\\
\qquad \qquad \qquad 237,238,239,240,241,242,243,245,247,250\}), \\
(\gamma =9,\ \beta =\{191,197,212,227,229,231,232,233,234,235,236,237,238,239,240,\\
\qquad \qquad \qquad 241,242,243,244,246,247\}). \\
\end{array}%
\end{equation*}

A suggestion for future work would be to consider groups other than $C_8$ or $D_8$ used in our main construction when performing the searches for self-dual codes. Another direction is to consider groups of larger lengths to obtain codes of greater lengths with weight enumerators not known in the literature yet.

\end{document}